\documentclass[preprint]{elsart}

\usepackage{multicol}
\usepackage{graphicx}
\usepackage{booktabs}
\usepackage{longtable}
\usepackage{amssymb,bm,mathrsfs,bbm,amscd}
\usepackage[tbtags]{amsmath}
\usepackage{lastpage}
\usepackage{CJK}
\usepackage{txfonts}
\usepackage{cases}
\journal{Chinese Physics C}
\begin{document}
\begin{CJK*}{GBK}{song}
\begin{frontmatter}

\title{Investigation of proton radioactivity with the effective liquid drop model }
\author[1]{SHENG Zong-Qiang \corauthref{cor}}
\corauth[cor]{Corresponding author} \ead{shengzongq@gmail.com}
\author[1]{SHU Liang-Ping}
\author[2]{FAN Guang-Wei}
\author[1]{MENG Ying}
\author[1]{QIAN Jian-Fa}

\address[1]{School of Science, Anhui University of Science and Technology, Huainan 232001, China}
\address[2]{School of Chemical Engineering, Anhui University of Science and Technology, Huainan 232001, China}
\begin{abstract}
  Proton radioactivity has been investigated
  using the effective liquid drop model with varying mass
  asymmetry shape and effective inertial coefficient. An effective nuclear radius
  constant formula replaces the old empirical one in the calculations.
  The theoretical half-lives are in good agreement with the available
  experimental data. All the deviations between the calculated logarithmic half-lives
  and the experimental values are less than 0.8.
  The root-mean-square deviation is 0.523. Predictions for the half-lives of
  proton radioactivity are made for elements across the periodic table. From the theoretical results, there are 11 candidate nuclei for proton radioactivity in the region $Z <51$. In the region $Z >83$, no nuclei are suggested as probable candidate nuclei for proton radioactivity within the selected range of half-lives studied.
\end{abstract}

\end{frontmatter}

\maketitle

\section{Introduction}
Nuclear decays, including alpha, beta and gamma decays, have been studied for more
than a century, employing many forms of theory and experiment \cite{Ren,Xu,Renyj}.
With the development of radioactive ion beams and related experimental facilities,
exotic nuclear physics has become an extremely interesting topic.
People can explore the unknown regions of the periodic table
 and look for many fundamentals in nuclear physics through studies of exotic nuclei \cite{Zhu,Jiang1,Jiang2,Cao,Zhou,Wang,Gan}. The investigation of exotic nuclei has led to the discovery of a new form of radioactivity \rule[2pt]{20pt}{0.5pt} proton radioactivity.

Proton radioactivity was observed for the first
time in the decay of an isomeric state of $^{53}$Co \cite{Jackson,Cerny}. Since
then, a number of proton-emitting nuclei have been
found in the region from Z = 51 to Z = 83 \cite{Woods,Sonzogni}.
These nuclei can emit protons from their ground states or low-lying isomeric states.
It is believed that more proton-emitting nuclei will be discovered in the future.

Proton radioactivity can be used as an excellent tool to extract rich nuclear structure information
such as shell structure, fine structure, wave function of the
parent nucleus, etc.\cite{Karny,Davids3,Arumugam}. Therefore, it is very important to learn more about proton radioactivity.

Many theoretical approaches and models have been employed to investigate proton
radioactivity \cite{Ferrei,Delion,Bhat,Dong1,Zhang1,Aggarwal,Routray,Ferreira,Routray2,Dong2,Wang2}.
There have also been many experiments for measuring proton
radioactivity \cite{Davids1,Davids2,Poli1,Poli2,Mahmud,Rykac,Woods2,Mukha,Robinson,Page2,Procter}.
In Refs. \cite{Goncalves,Duarte}, Gon\c{c}alves
 and Duarte proposed the effective liquid drop
model to describe exotic decays. Two kinds of shape parametrization mode and
two kinds of inertial coefficient are introduced in this model. Strikingly, one can reproduce the data of alpha and exotic decays by using
only two parameters \rule[2pt]{20pt}{0.5pt} the nuclear radius constant $r_{0}$ and
the preexponential factor $\lambda_{0}$. In our previous work, we have investigated cluster radioactivity using this model and got excellent results \cite{Sheng}.

In this work, we will investigate proton radioactivity by using the effective
liquid drop model with varying mass asymmetry shape and effective inertial coefficient.
In view of the importance of $r_{0}$  for this model, an effective nuclear radius constant
formula will be introduced instead of the empirical one. Systematic calculations of the proton radioactivity will be performed across the periodic table.

This paper is organized as follows. In Sec. 2, a brief review of the theoretical framework
is provided. In Sec. 3, numerical results and discussions are presented. A summary is given in Sec. 4.

\section{Theoretical framework}
In this section, we give a brief theoretical framework. One can
find the details of the model in Refs. \cite{Goncalves,Duarte}. The
decaying system can be regarded as two intersectant spherical
fragments with different radii. The shape configuration and
geometric parameters are shown in Figure 1.

\begin{figure}[htb]
\begin{center}
\includegraphics[width=6cm]{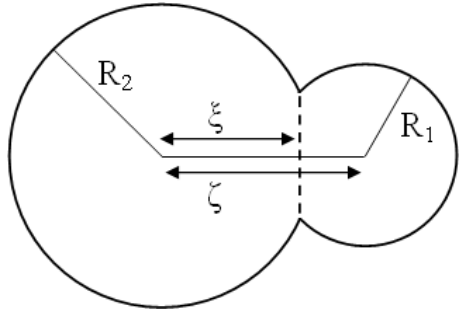}
\end{center}
\caption{Shape parametrization of dinuclear system. The two
spherical segments radii are $R_{1}$ and $R_{2}$ respectively. The
distance between the geometrical centers of the two spherical
fragments is marked $\zeta$. $\xi$ is the distance from the plane of
the intersection to the geometrical center of the heavier fragment.}\label{fig1}
\end{figure}

There are four geometric shape parameters
(${R_{1}, R_{2}, \zeta, \xi}$) used to describe the dinuclear system. Three constraint relations used in the
model are given in the following three equations. First,
\begin{equation}
2(R\,_{1}^{3}+R\,_{2}^{3})+3[R\,_{1}^{2}(\zeta-\xi)+R\,_{2}^{2}\,\xi]-[(\zeta-\xi)^{3}+\xi^{3}]-4R\,_{m}^{3}=0,
\end{equation}
where $R_{m}$ denotes the radius of the parent nucleus. Second,
\begin{equation}
R\,_{1}^{2}-R\,_{2}^{2}-(\zeta-\xi)^{2}+\xi^{2}=0.
\end{equation}
Third, at the end of the prescission phase, the system will reach a
critical state. At this time, the radii of the two spherical fragments
are denoted as $\overline{R}_{1}$ and $\overline{R}_{2}$ for the
cluster and the daughter nucleus, respectively. In the mode of varying mass
asymmetry (VMAS), the radius of the light fragment is fixed as
\begin{equation}
R_{1}-\overline{R}_{1}=0.
\end{equation}

In such a picture, the problem can conveniently be reduced to the
one-dimensional barrier penetrability problem, and the quantum transition
rate can be similarly calculated in accordance with the Gamow
alpha decay theory \cite{Gam}. The expression of the barrier
penetrability factor is written as
\begin{equation}
\mathcal
{P}=exp\{-\frac{2}{\hbar}\int_{\zeta_{0}}^{\zeta_{C}}\sqrt{2\mu_{\textrm{eff}}(V-Q)}d\zeta\},
\end{equation}
where $\mu_{\textrm{eff}}$ is the effective inertial coefficient defined in Ref. \cite{Duarte}. $\zeta_{0}$ and $\zeta_{C}$ are respectively the inner and outer
turning points. $\zeta_{0}$ is written as $\zeta_{0}=R_{m}-\overline{R}_{1}$.
In view of the importance of angular momentum for proton radioactivity, $\zeta_{C}$ is written as

\begin{equation}
\zeta_{C}=\frac{Z_{1}Z_{2}e^{2}}{2Q}+\sqrt{\left( \frac{Z_{1}Z_{2}e^{2}}{2Q} \right)^{2}
+\frac{l(l+1)\hbar^{2}}{2\mu_{\textrm{eff}}\,Q}}.
\end{equation}
Q is the decay energy and V is the total
one-dimensional effective liquid drop potential. The decay energy Q
is calculated by the relation $Q=M-M_{1}-M_{2}$, where the masses in
the Q value expression are extracted from the latest atomic mass table \cite{Wang1}. For details of the meanings and
the expression of V, please see Ref. \cite{Goncalves}.

The final radii of the fragments are written as
\begin{equation}
\overline{R}_{i}=\left(\frac{Z_{i}}{Z_{m}}\right)^{1/3}R_{m}.
\end{equation}
$Z_{m}$ and $R_{m}$ are the proton number and radius of the parent
nucleus. $Z_{i}$ is the proton number of the fragment.

As mentioned in the introduction, the nuclear radius is of great
importance in the present calculation. Previously, a
constant $r_{0}$ was always employed in the calculation. However, it is well
known from available experimental data that the nuclear radius
constant $r_{0}$ is far from being constant. For this reason, the radius of the parent nucleus is determined by a more
precise formula which includes the isospin-dependence:
\begin{equation}
R_{m}=r_{0}A_{m}^{1/3}=\frac{1.38}{1.20}\times(1.2332+\frac{2.8961}{A_{m}}-0.18688\times
I)A_{m}^{1/3},
\end{equation}
where \textit{I} is the relative neutron excess
$I=(N_{m}-Z_{m})/A_{m}$. In the present work, we will use this
formula to replace the empirical
$R_{m}=1.38 A_{m}^{1/3}$.

The decay rate is calculated by
\begin{equation}
\lambda=\lambda_{0}\mathcal {P},
\end{equation}
where $\lambda_{0}$ is the preexponential factor. The value of
$\lambda_{0}$ is written as
 \cite{Duarte2}:
\begin{equation}
\lambda_{0}=0.5\times10^{21}s^{-1}.
\end{equation}
With $\lambda_{0}$ fixed, the half-life can be
calculated by
\begin{equation}
T=\ln2/\lambda.
\end{equation}

\section{Numerical results and discussion}
The calculated half-lives of proton radioactivity are
listed in Table 1 and the available experimental data are also included for
comparison. We only list the results for the most probable
proton decay (ground-state to ground-state transitions).
For cluster radioactivity,
the angular momentum carried by the cluster is not very
large ($l \leq 6$) and the centrifugal barrier is much lower than
the Coulomb barrier. The contribution of the angular momentum can be
ignored due to the relatively large reduced mass of
the cluster-core system \cite{Duarte}.
For proton radioactivity, however, the contribution of the
angular momentum is very important.
The centrifugal barrier has an evident effect on proton radioactivity.
In the calculations, therefore, we take into account the contribution of the
angular momentum on the half-life of proton radioactivity.

\begin{table}[!htb]
\centering \caption{The theoretical half-lives of proton radioactivity.
The experimental Q values are extracted from Ref. \cite{Wang1}.
All the experimental half-lives and \textit{l} values are
from Ref. \cite{Sonzogni}. Q value is in MeV, and half-life T in seconds.
}\label{tab1}\vspace{0.5cm}
\begin{tabular*}{120mm}{c l c c c c r r}
\hline
No.&Nucleus& Z& $l(\hbar)$ & Q&$ \mathcal
{P}$&$log_{10}T_{Theo.}$ & $log_{10}T_{Exp.}$\\
\hline
1&$^{105}$Sb&51&2&0.483&8.91$\times$ 10$^{-24}$&1.830&2.049\\
2&$^{109}$I &53&2&0.820&2.35$\times$ 10$^{-17}$&$-$4.592&$-$3.987\\
3&$^{112}$Cs&55&2&0.814&3.72$\times$ 10$^{-18}$&$-$3.791&$-$3.301\\
4&$^{113}$Cs&55&2&0.974&5.61$\times$ 10$^{-16}$&$-$5.570&$-$4.777\\
5&$^{145}$Tm&69&5&1.740&9.44$\times$ 10$^{-16}$&$-$6.195&$-$5.409\\
6&$^{147}$Tm&69&5&1.058&5.39$\times$ 10$^{-22}$&0.049&0.591\\
7&$^{150}$Lu&71&5&1.270&4.83$\times$ 10$^{-20}$&$-$1.905&$-$1.180\\
8&$^{151}$Lu&71&5&1.241&2.41$\times$ 10$^{-20}$&$-$1.602&$-$0.896\\
9&$^{155}$Ta&73&5&1.776&2.49$\times$ 10$^{-16}$&$-$5.616&$-$4.921\\
10&$^{156}$Ta&73&2&1.014&3.98$\times$ 10$^{-21}$&$-$0.810&$-$0.620\\
11&$^{157}$Ta&73&0&0.935&1.25$\times$ 10$^{-21}$&$-$0.318&$-$0.523\\
12&$^{160}$Re&75&2&1.278&2.25$\times$ 10$^{-18}$&$-$3.572&$-$3.046\\
13&$^{161}$Re&75&0&1.197&1.56$\times$ 10$^{-18}$&$-$3.414&$-$3.432\\
14&$^{164}$Ir&77&5&1.570&1.06$\times$ 10$^{-18}$&$-$3.245&$-$3.959\\
15&$^{166}$Ir&77&2&1.152&3.90$\times$ 10$^{-21}$&$-$1.153&$-$0.824\\
16&$^{167}$Ir&77&0&1.071&8.30$\times$ 10$^{-21}$&$-$1.140&$-$0.959\\
17&$^{171}$Au&79&0&1.452&6.29$\times$ 10$^{-17}$&$-$5.019&$-$4.770\\
18&$^{177}$Tl&81&0&1.162&9.45$\times$ 10$^{-21}$&$-$1.196&$-$1.174\\
19&$^{185}$Bi&83&0&1.543&8.76$\times$ 10$^{-18}$&$-$4.875&$-$4.229\\
\hline
\end{tabular*}
\end{table}

In Table 1, the first three columns list the sequence numbers, parent nuclei
 and corresponding proton numbers, respectively.
The fourth column is the experimental \textit{l} value \cite{Sonzogni}. The
fifth column denotes the experimental Q values extracted from
 the latest atomic mass table \cite{Wang1}. The sixth column is the theoretical penetrability factor $\mathcal
{P}$.
The seventh and eighth columns list the calculated logarithmic half-lives and
the experimental values \cite{Sonzogni}, respectively.

From Table 1, one can see that in most cases the calculated half-lives are in good agreement
with the experimental values. All the deviations between
 the calculated logarithmic half-lives
 and the experimental data are less than 0.8. The biggest and smallest
 deviations are 0.793 (in $^{113}$Cs) and 0.018 (in $^{161}$Re), respectively.

The root-mean-square (rms) deviation between the calculated
logarithmic half-lives and the experimental ones is defined as:
\begin{equation}
\sigma=\Big[\sum_{i=1}^{19}(log_{10}T_{Theo.}-log_{10}T_{Exp.})^{2}
/19\Big]^{1/2}=0.523.
\end{equation}
The rms deviation is small, showing that the present calculations
for proton radioactivity are reliable.

For a clear insight into the reliability of the present
calculations, we will discuss the theoretical results graphically.
The deviations between the calculated logarithmic half-lives and
the experimental data are shown in Figure 2.

\begin{figure}[htb]
\begin{center}
\includegraphics[width=12cm]{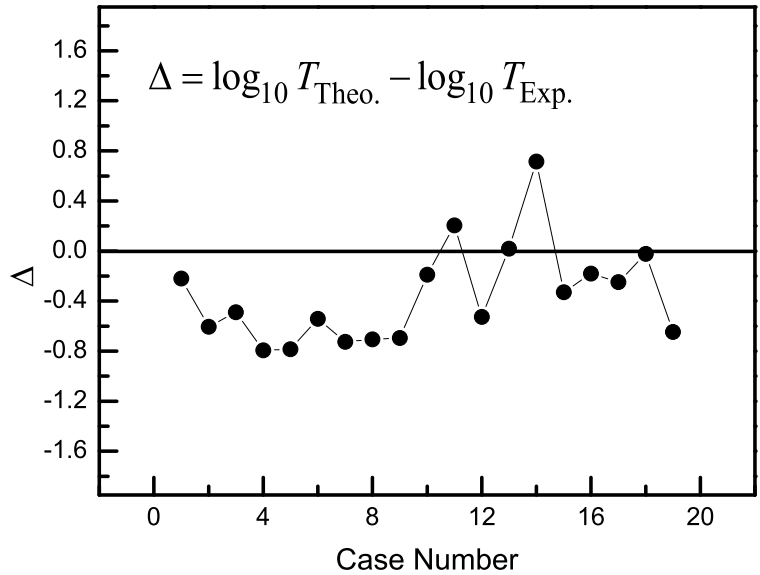}
\end{center}
\caption{Deviations between the calculated logarithmic
half-lives and experimental data. The deviation is defined as: $\Delta=log_{10}T_{Theo.}-log_{10}T_{Exp.}$.}\label{fig2}
\end{figure}

One can clearly see from Figure 2 that the theoretical results are
close to the experimental data, with all deviations between the
calculated logarithmic half-lives and the experimental data
less than 0.8.

The theoretical half-lives show a strong dependence on the
 orbital angular momentum of the emitted proton. The \textit{l} values in Table 1 are the  values suggested in the experimental literature. It should be pointed out that in all
the experiments performed to date, only the half-life and the
energy of the proton are measured. The spin and parity are not
experimentally observed quantities - the values given in experimental
papers are based on the calculated decay rates. The theoretical results can get better agreement with the experimental data if a suitable \textit{l} value is employed in the present model.

 In Ref. \cite{Robinson}, ground-state proton radioactivity has been identified from $^{121}$Pr. A transition with a half-life $T_{1/2}=10^{+6}_{-3}$ ms ($log_{10}T=-2.000$) has been observed and is assigned to the decay of a highly prolate deformed $3/2^{+}$ or $3/2^{-}$ Nilsson state. In Figure 3 we show the variation of the half-life of proton emission for $^{121}$Pr as a function of angular momentum.

\begin{figure}[htb]
\begin{center}
\includegraphics[width=12cm]{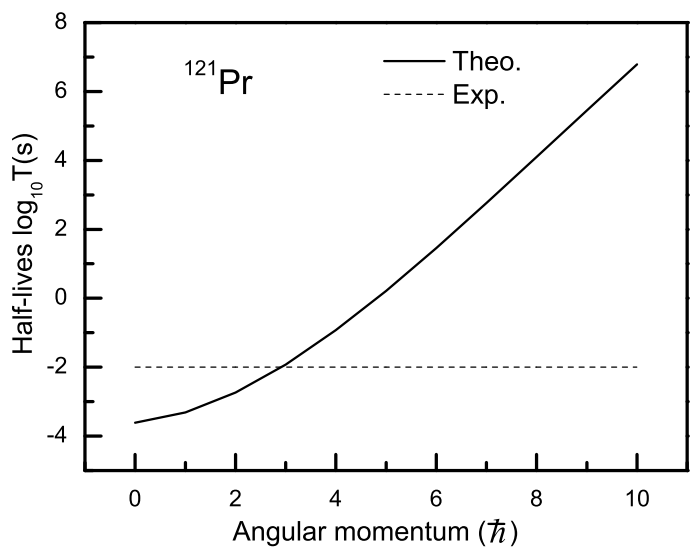}
\end{center}
\caption{Angular momentum dependence of the half-life of proton radioactivity for $^{121}$Pr.}\label{fig3}
\end{figure}

The solid line denotes the calculated results. It can be seen that the half-life values have a strong dependence on the orbital angular momentum of the emitted proton. There is an increase of 10 orders of magnitude in the half-life values when the orbital angular momentum varies from 0 to 10$\hbar$. The measured half-life (dashed line in Figure 3) is $log_{10}T=-2.000$. If the orbital angular momentum is chosen as $l=3$, the calculated value is in excellent agreement with the experimental one.

 In Ref. \cite{Davids2}, proton radioactivity from $^{141}$Ho and $^{131}$Eu has been identified. The $^{141}$Ho proton transition has a half-life $T_{1/2}=4.2(4)$ ms ($log_{10}T=-2.376$), and is assigned to the decay of the $7/2^{-}[523]$
Nilsson state. The $^{131}$Eu transition has a half-life $T_{1/2}=26(6)$ ms ($log_{10}T=-1.585$),
consistent with the decay from either the $3/2^{+}[411]$ or $5/2^{+}[413]$ Nilsson orbital. In the present model, the calculated
logarithmic half-life is $log_{10}T=-1.948$ for $^{141}$Ho if the orbital angular momentum is chosen as $l=5$, and the calculated value for $^{131}$Eu is $log_{10}T=-1.186$ if $l=4$. The calculated results are in good agreement with the experimental ones.

From the above discussions, one can see that the present model is reliable
for calculating the half-lives of proton radioactivity. This gives
us confidence to perform theoretical predictions for the possible
candidate nuclei for proton radioactivity.

We perform a large number of systematic
calculations for half-lives of proton radioactivity throughout the periodic table.
From Table 1, the experimental transferred angular momenta are usually $l= 0,\,\, 2, \,\,5$. Because the angular momenta carried by the emitted proton of the probable candidate nuclei for proton radioactivity are unknown, the half-lives are calculated under the assumption of the three probable transferred angular momenta ($l= 0,\,\, 2, \,\,5$).

From the calculated results, we only select candidate nuclei with $-10<log_{10}T_{Theo.}<10$ to display. The predicted probable
candidate nuclei and their
half-lives for proton radioactivity are listed in Table 2.

\begin{longtable}{c c c r c r }
\caption[!htp]{ Predicted probable
candidate nuclei for proton radioactivity and their calculated half-lives under the three probable transferred angular momenta ($l= 0,\,\, 2, \,\,5$). The experimental Q values are extracted from Ref. \cite{Wang1}.
 Q value is in MeV, and half-life T is in seconds. }
\label{tab2}\vspace{0.5cm}\\
\hline\hline $\begin{array}{c}$$\\Nucleus$$
\end{array}$ & $\begin{array}{c} $$\\Z$$
\end{array}$ & $\begin{array}{c} $$\\Q$$
\end{array}$ & $\begin{array}{c} $$\\l=0$$
\end{array}$ & $\begin{array}{c} $$log_{10}T_{Theo.}\\l=2$$
\end{array}$ & $\begin{array}{c} $$\\l=5$$
\end{array}$ \\
\hline
\endfirsthead
\multicolumn{6}{c}%
{{\bfseries \tablename\ \thetable{} -- continued from previous page}} \\
\hline\hline  $\begin{array}{c}$$\\Nucleus$$
\end{array}$ & $\begin{array}{c} $$\\Z$$
\end{array}$ & $\begin{array}{c} $$\\Q$$
\end{array}$ & $\begin{array}{c} $$\\l=0$$
\end{array}$ & $\begin{array}{c} $$log_{10}T_{Theo.}\\l=2$$
\end{array}$ & $\begin{array}{c} $$\\l=5$$
\end{array}$ \\
\hline
\endhead
\hline \multicolumn{6}{r}{{(Continued on next page)}} \\
\endfoot
\hline
\endlastfoot
\hline
$^{42}$V&23&0.242&$-$7.328&$-$5.824&$-$1.528\\
$^{50}$Co&27&0.098&8.454&9.847&14.449\\
$^{54}$Cu&29&0.387&$-$7.997&$-$6.678&$-$2.806\\
$^{55}$Cu&29&0.298&$-$5.325&$-$4.003&0.024\\
$^{68}$Br&35&0.560&$-$8.571&$-$7.388& $-$3.839\\
$^{69}$Br&35&0.450&$-$6.344&$-$5.160& $-$1.498\\
$^{73}$Rb&37&0.600&$-$8.373&$-$7.227&$-$3.749\\
$^{76}$Y&39&0.629&$-$7.988&$-$6.874&$-$3.449\\
$^{81}$Nb&41&0.750&$-$8.960&$-$7.879&$-$4.574\\
$^{85}$Tc&43&0.852&$-$9.475&$-$8.423&$-$5.206\\
$^{89}$Rh&45&0.700&$-$6.696&$-$5.670&$-$2.399\\
$^{104}$Sb&51&0.509&0.078&1.028&4.269\\
$^{108}$I&53&0.600&$-$1.394&$-$0.464& 2.691\\
$^{118}$La&57&0.378&8.407&9.293&12.445\\
$^{122}$Pr&59&0.526&3.448&4.318&7.371\\
$^{126}$Pm&61&0.759&$-$1.492&$-$0.637& 2.305\\
$^{127}$Pm&61&0.545&3.795&4.647&7.650\\
$^{130}$Eu&63&1.028&$-$4.974&$-$4.132& $-$1.293\\
$^{133}$Eu&63&0.675&1.101&1.937&4.860\\
$^{136}$Tb&65&0.918&$-$2.784&$-$1.960&0.870\\
$^{137}$Tb&65&0.759&0.047&0.869&3.735\\
$^{142}$Ho&67&0.554&6.340&7.143&10.020\\
$^{148}$Tm&69&0.489&9.798&10.585&13.436\\
$^{152}$Lu&71&0.833&0.827&1.605&4.350\\
$^{153}$Lu&71&0.609&6.384&7.159& 9.946\\
$^{162}$Re&75&0.764&3.831&4.582&7.272\\
$^{163}$Re&75&0.706&5.278&6.027&8.726\\
$^{169}$Ir&77&0.621&8.675&9.411&12.088\\
$^{169}$Au&79&1.961&$-$8.808&$-$8.069&$-$5.588\\
$^{170}$Au&79&1.474&$-$5.221&$-$4.485&$-$1.949\\
$^{172}$Au&79&0.900&2.411& 3.140&5.747\\
$^{173}$Au&79&0.992&0.730&1.459&4.053\\
$^{176}$Tl&81&1.250&$-$2.320&$-$1.598&0.937\\
$^{178}$Tl&81&0.738&6.866& 7.581&10.178\\
$^{179}$Tl&81&0.727&7.147&7.861&10.458\\
$^{184}$Bi&83&1.327&$-$2.675&$-$1.964&0.533\\
$^{186}$Bi&83&1.083&0.546&1.253&3.775\\
$^{187}$Bi&83&1.019&1.579&2.285&4.813\\
\hline
\end{longtable}

At present, theoretical and experimental studies on proton radioactivity are mainly focused on the region $51\leq Z \leq 83$. In Table 2, we are surprised to find that there are 11 candidate nuclei for proton radioactivity in the $Z <51$ region, which is a very interesting result. In the region $Z >83$, we don't find any candidate nuclei for proton radioactivity within our selected range of half-lives ($-10<log_{10}T_{Theo.}<10$).

\section{Summary}
In this work, proton radioactivity has been investigated
using the effective liquid drop model with varying mass
asymmetry shape and effective inertial coefficient. In view of the
importance of $r_{0}$ for this model, a new effective nuclear radius
constant formula replaces the old empirical one.
 The theoretical half-lives are in good agreement with the available
 experimental data. All the deviations between the calculated logarithmic half-lives
 and the experimental data are less than 0.8.
 The root-mean-square deviation between the calculated
logarithmic half-lives and the experimental ones is 0.523. For proton radioactivity from $^{121}$Pr, $^{141}$Ho and $^{131}$Eu, the calculated half-lives are in good agreement with the experimental values if we select suitable $l$ values in the present model. We make some predictions for half-lives of
 proton radioactivity throughout the periodic table. We find that there are 11 candidate nuclei for proton radioactivity in the region $Z <51$. In the region $Z >83$, we don't find any candidate nuclei for proton radioactivity within our selected half-life range. Our calculated results may be useful for future experiments.

\begin{center}
{\large Acknowledgments }
\end{center}
This work is supported by Supported by National Natural Science Foundation of China under Grant No. 11247001, by the Natural Science Foundation of the Higher Education Institutions of Anhui Province, China under Grant Nos. KJ2012A083 and KJ2013Z066, and by Anhui Provincial Natural Science Foundation under Grant No. 1408085MA05.

\clearpage
\end{CJK*}
\end{document}